\documentclass[a4paper, 11pt]{article}

\usepackage{amsmath, amssymb, cite, indentfirst}
\usepackage[dvips]{graphicx}

\title{Numerical Approach to Central Limit Theorem for Bifurcation Ratio
of Random Binary Tree}
\author{Ken Yamamoto and Yoshihiro Yamazaki}
\date{\it \small 
Department of Physics, Waseda University, Tokyo, 169-8555, Japan}

\usepackage{indentfirst}

\newcommand{\oper}[2]{(#1\ #2)}
\newcommand{\domain}{\mathcal{D}}

\DeclareMathOperator{\erf}{erf}

\begin{document}
\maketitle

\begin{abstract}
A central limit theorem for binary tree is numerically examined.
Two types of central limit theorem for higher-order branches
are formulated.
A topological structure of a binary tree is expressed
by a binary sequence,
and the Horton-Strahler indices are calculated
by using the sequence.
By fitting the Gaussian distribution function to our numerical data,
the values of variances are determined
and written in simple forms.
\end{abstract}

\section{Introduction}
Branching patterns are widely spread in the nature \cite{Ball, Fleury}.
Some patterns appear to be quite similar
to each other even if their generation process are different.
The branching patterns are characterized from various standpoints.
For example,
a property related to spatial configurations is called geometric,
including length, spatial symmetry, and fractality.
On the other hand,
a property based on graph-theoretic structure
(and not on spatial extent) is called topological.
Connectivity and degree distributions of complex networks
are typical and important topological structures.
In particular,
the topological structure of a branching pattern can be expressed by
a binary-tree graph.

\begin{figure}[htb]
\centering
\includegraphics[width = 0.4\textwidth]{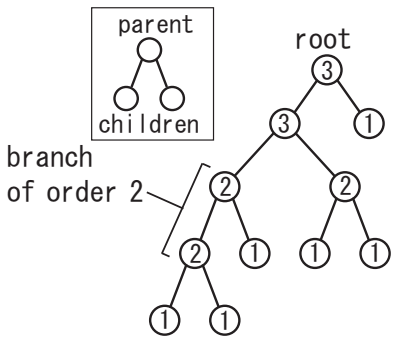}
\caption{
	An example of a binary tree of magnitude 6.
	The numbers on the nodes represent the Horton-Strahler indices.
}
\label{fig:binary}
\end{figure}

A {\it full binary tree} is a tree graph
(i.e., a connected graph without loops) 
where
every node has exactly zero or two `children'
(see Fig. \ref{fig:binary} for reference).
For simplicity,
we use the term `binary tree' instead of `full binary tree' hereafter,
since we focus on only full binary trees throughout the paper.
A node without any children is called {\it leaf},
the node without `parents' is called {\it root},
and the number of leaves is called {\it magnitude}.
Binary trees have been mainly investigated in computer science,
and
frequently used in order to represent some types of data structures such
as binary search tree, binary heap, and expression tree \cite{Wirth, Aho}.

In order to derive topological characteristics of branching patterns,
a method of branch ordering has been introduced
by Horton \cite{Horton} and Strahler \cite{Strahler}.
With this method, ramification complexity and a hierarchical structure
of branching patterns can be measured.
For each node $v$ in a binary tree $T$,
the {\it Horton-Strahler index} $S(v)$ is defined recursively as
\begin{equation}
S(v)=
\begin{cases}
1, & \mbox{if } v \mbox{ is a leaf,}\\
\max\{S(v_1),S(v_2)\}+\delta_{S(v_1),S(v_2)},
				& \mbox{if } v_1 \mbox{ and } v_2
				\mbox{ are the children of } v,\\
\end{cases}
\label{eq:horton}
\end{equation}
where $\delta_{i,j}$ is the Kronecker delta.
We define a {\it branch} of order $r$ as a maximal path
connecting nodes of order $r$.
The ratio of the number of branches of two subsequent orders
is called the {\it bifurcation ratio},
and it has been found in many branching patterns that
the bifurcation ratio takes almost constant value for different orders,
which is known as ``Horton's law of stream numbers''
especially in river networks \cite{Horton}.
Horton-Strahler analysis has been applied to
a wide range of branching patterns
\cite{Berry, Gabneshaiah, Berntson, Feder, Ossadnik, Horsfield,
	Viennot_CG, Vannimenus, Zaliapin}.

A simple model called {\it random model} or {\it equiprobable model},
formulated by Shreve \cite{Shreve},
is a finite probability space $(\Omega_n, P_n)$,
where $\Omega_n$ denotes the sample space consisting of
topologically distinct binary trees of magnitude $n$,
and $P_n$ is the uniform probability measure on $\Omega_n$.
We also introduce a random variable $S_{r,n}:\Omega_n\to\mathbb{N}\cup\{0\}$
such that $S_{r,n}(T)$ represents the number of branches of order $r$
in a binary tree $T\in\Omega_n$.
Horton's law on $(\Omega_n, P_n)$ is stated in the form
\begin{equation}
\frac{E(S_{r,n})}{E(S_{r-1,n})}\to\frac{1}{4}
\quad \mbox{as } n\to\infty,
\label{eq:Horton}
\end{equation}
where $E(\cdot)$ denotes the average on $(\Omega_n, P_n)$,
and $r=2,3,\cdots$.
Analytical or combinatorial properties of $S_{r,n}$ are
discussed in
\cite{Werner, Meir, Gupta, Devroye, Prodinger, Toroczkai, Yamamoto}
for example.

Wang and Waymire analytically proved the central limit theorem
\begin{equation}
\sqrt{n}\left(\frac{S_{2,n}}{n}-\frac{1}{4}\right)
\Rightarrow N\left(0,\frac{1}{16}\right)
\quad \mbox{as } n\to\infty,
\label{eq:central}
\end{equation}
where ``$\Rightarrow$'' denotes convergence in distribution,
and $N(\mu, \sigma^2)$ denotes Gaussian distribution
with mean $\mu$ and variance $\sigma^2$ \cite{Wang}.
Eq. \eqref{eq:central} is equivalently expressed as
\[
P_n\left(\sqrt{n}\left(\frac{S_{2,n}}{n}-\frac{1}{4}\right)\le x\right)
\rightarrow\frac{4}{\sqrt{2\pi}}\int_{-\infty}^x
e^{-8t^2}\,{\mathrm d}t \quad \mbox{as } n\to\infty.
\]
In the same way as Eq. \eqref{eq:Horton},
we expect the following relations
\[
E\left(\frac{S_{r,n}}{S_{r-1,n}}\right)\to \frac{1}{4},\quad
E\left(\frac{S_{r,n}}{n}\right)\to \frac{1}{4^{r-1}}
\quad\mbox{as } n\to\infty.
\]
And, Eq. \eqref{eq:central} is considered to be naturally generalized to
\begin{subequations}
\label{eq:general}
\begin{eqnarray}
\sqrt{n}\left(\frac{S_{r,n}}{S_{r-1,n}}-\frac{1}{4}\right)
&\Rightarrow& N(0, \sigma_r^2), \label{eq:general-1}\\
\sqrt{n}\left(\frac{S_{r,n}}{n}-\frac{1}{4^{r-1}}\right)
&\Rightarrow& N(0, \tilde{\sigma}_r^2), \label{eq:general-2}
\end{eqnarray}
\end{subequations}
where $\sigma_r^2$ and $\tilde{\sigma}_r^2$ are
variances depending on the order $r$.
However, the proof of Eqs. \eqref{eq:general}
has not been performed analytically or numerically so far,
and the values of $\sigma_r$ and $\tilde{\sigma}_r$
have not been obtained for $r\ge 3$.
In the present paper,
we propose a method of calculating Horton-Strahler indices of a binary tree
by using binary sequence,
and show numerical evidence for the validity of Eqs. \eqref{eq:general}.

\section{Correspondence between Binary Trees and Dyck Paths}
\label{sec:correspond}

\begin{figure}
\centering
\includegraphics[width = 0.5\textwidth]{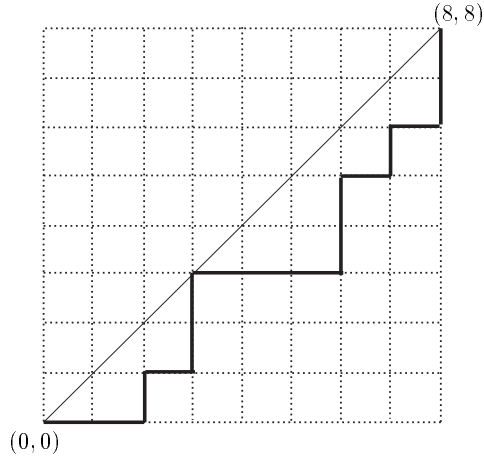}
\caption{An example of a Dyck path of length 16.
Dashed lines indicate grid lines of $\mathbb{Z}^2$.
All the Dyck paths lie below the diagonal line}
\label{fig:Dyck}
\end{figure}

A {\it Dyck path} of length $2(n-1)$ is a sequence of points
$(s_0,\cdots,s_{2(n-1)})$ on a two-dimensional lattice $\mathbb{Z}^2$
from $s_0=(0,0)$ to $s_{2(n-1)}=(n-1, n-1)$
such that
each point $s_i=(x_i, y_i)$ satisfies $x_i\ge y_i$ and
each elementary step $(s_i,s_{i+1})$ is either rightward or upward
(see Fig. \ref{fig:Dyck}).

For each Dyck path,
a binary sequence of length $2(n-1)$ is generated
by replacing a rightward step with `1' and an upward step with `0'.
The binary sequences generated by this replacement are formally called
{\it Dyck words on the alphabet $\{1,0\}$} \cite{Duchon},
and for simplicity we call them {\it Dyck sequences}
throughout the paper.
Clearly, Dyck sequences share the two properties:
(i) the total number of `0' (and also `1') is $n-1$,
(ii) cumulative number of `0' is never greater than that of `1'.

\begin{figure}[htbp]
\centering
\includegraphics[width = 0.8\textwidth]{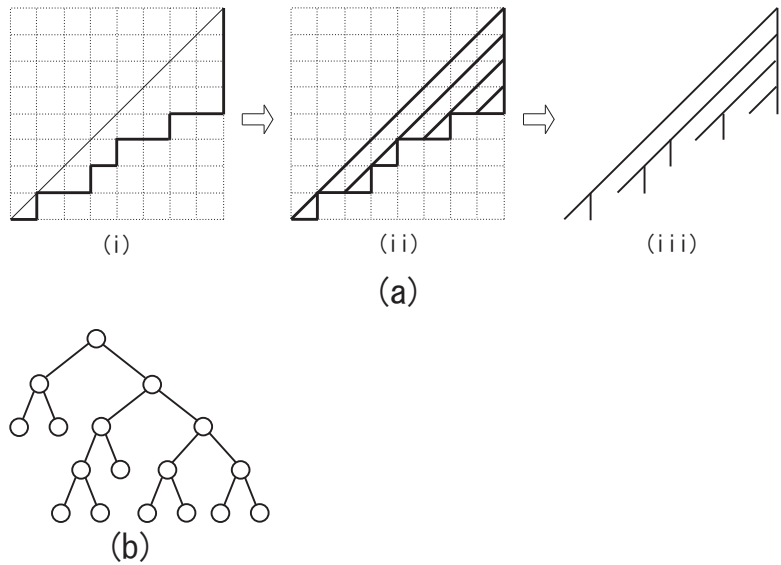}
\caption{
	(a) An illustration of how to get a binary tree from a Dyck path.
	(i) The initial Dyck path of length 16.
	(ii) The Dyck path with diagonals from upper right to lower left.
	(iii) The diagonals and vertical steps.
		The structure of a binary tree can be seen.
	(b) The binary tree corresponding to (a-iii).
}
\label{fig:correspondence1}
\end{figure}

A correspondence between
the Dyck paths of length $2(n-1)$ and the binary trees of magnitude $n$
is explained as follows (see Fig. \ref{fig:correspondence1} for reference).
(i) Start with a Dyck path of length $2(n-1)$.
(ii) Draw diagonal lines from upper right to lower left
	which are never below the Dyck path.
(iii) Extract only the diagonals and the vertical lines in the Dyck path.
It is found that the pattern obtained from this process is
topologically the same as a binary tree
of magnitude $n$,
shown in Fig. \ref{fig:correspondence1} (b).
Note that
each Dyck path has one-to-one correspondence to a binary tree.
Therefore, a Dyck path possesses
the same topological structure as the corresponding binary tree.

\begin{figure}[t!]
\centering
\includegraphics[width = 0.7\textwidth]{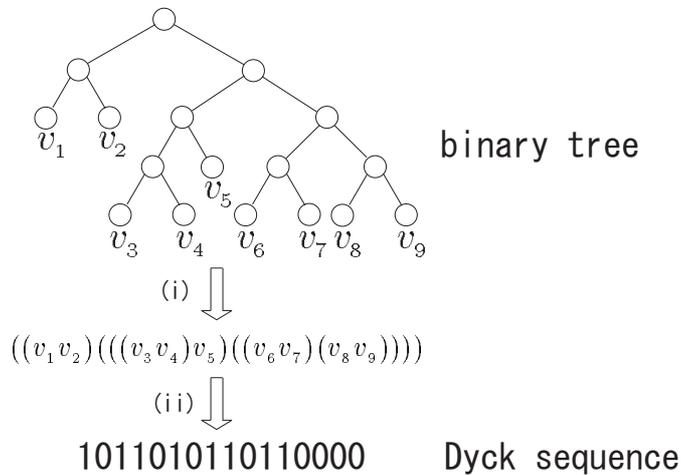}
\caption{An illustration of correspondence
between a binary tree of magnitude 9 and a Dyck sequence of length 16.
In the step (i), a binary tree is converted into a sequence consisting of
$v_1,\cdots,v_9$ and `$\oper{}{}$'.
In the step (ii), a Dyck sequence is generated by the rule of replacement.}
\label{fig:correspondence}
\end{figure}

The above method can be reformulated in a different way,
where a Dyck sequence is generated from a binary tree.
Here, a binary tree is regarded as a graph representing
a successive merging process of two adjacent nodes,
and each merging is expressed by putting two nodes
in parentheses `$\oper{}{}$'.
Thus, the topological structure of a binary tree $T\in\Omega_n$ is
fully expressed by a sequence of the leaves $v_1,\cdots,v_n$ of $T$ and 
$n-1$ pairs of `$\oper{}{}$'
[an example is shown as step (i) in Fig. \ref{fig:correspondence}].
A correspondence between a binary tree $T\in\Omega_n$
and a Dyck sequence of length $2(n-1)$
consists of the following two steps.
(i) Convert $T$ into a sequence of $v_1,\cdots,v_n$ and `$\oper{}{}$'.
(ii) Eliminate `$v_1$' and `$($',
and replace $v_2,\cdots,v_n$ with `$1$' and `$)$' with `$0$.'
A generated binary sequence proves to be a Dyck sequence
and the correspondence is one-to-one.
Fig. \ref{fig:correspondence} illustrates this correspondence.
Note that
this process is similar to an expression tree and reverse Polish notation
in formula manipulation \cite{Koshy}.

\begin{figure}
\centering
\includegraphics[width = \textwidth]{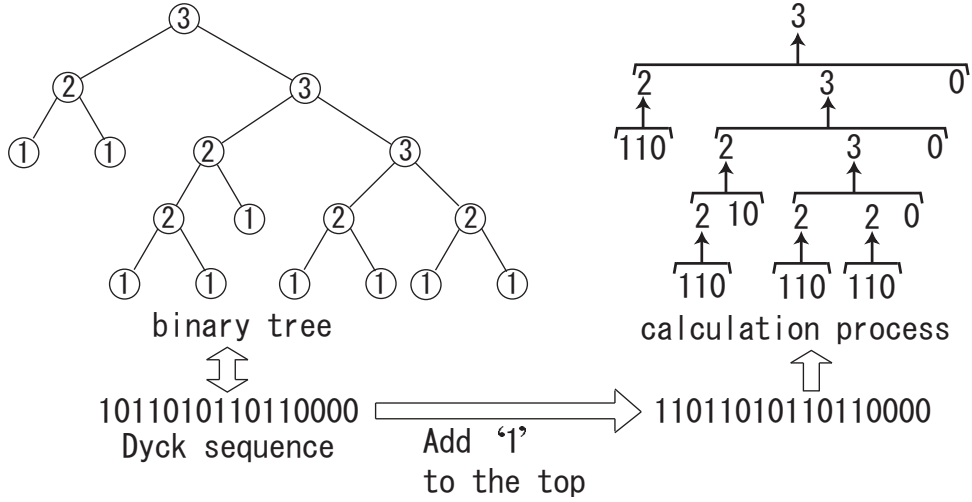}
\caption{Similarity between a structure of the Horton-Strahler indices and
corresponding calculation process.}
\label{fig:calculate}
\end{figure}

The Horton-Strahler indices of a binary tree
can be calculated through the corresponding Dyck sequence.
The method consists of the following two steps:
(i) Add `1' to the top of the Dyck sequence.
(ii) Replace a segment `$m\; n\; 0$' ($m,n>0$)
with a single number `$\max\{m,n\}+\delta_{m,n}$' recursively
until the length of a sequence becomes 1.
It is found that
the number of times of a transformation
`$(r-1)\; (r-1)\; 0$' $\to$ `$r$' is identical with $S_{r,n}(T)$ for $r\ge2$.
Note that the operation (ii) is similar to Eq. \eqref{eq:horton}
as shown in Fig. \ref{fig:calculate}.

\section{Generation of Random Dyck Paths}
\label{sec:generation}

A basic method for generation of random Dyck paths is
summarized in \cite{Alonso}.
In this section,
we present a method
in a little different manner from \cite{Alonso}.
We also propose a graphical representation for the generation process.

Let $\domain$ denote the set of points in $\mathbb{Z}^2$
where at least one Dyck path passes,
that is,
$\domain\equiv\{(x,y)\in\mathbb{Z}^2\mid 0\le x,y\le n-1,\ x\ge y\}$.
We assign `transition probabilities' $P_{\rightarrow}(x,y)$ and
$P_{\uparrow}(x,y)$ on each point $(x,y)\in\domain$.
Each elementary step $(s_i, s_{i+1})$
of a Dyck path $(s_1,\cdots,s_{2(n-1)})$
is selected stochastically:
stepping rightward with a probability $P_{\rightarrow}(s_i)$
and upward with $P_{\uparrow}(s_i)$.
A set of transition probabilities yields
a generation probability of a Dyck path $(s_0,\cdots,s_{2(n-1)})$,
which is given by
\[
P(s_0,\cdots,s_{2(n-1)})=\prod_{i=0}^{2(n-1)-1}p_i,\quad
\mbox{where }
p_i = 
\begin{cases}
P_{\rightarrow}(s_i), & \mbox{if $(s_i,s_{i+1})$ is rightward}, \\
P_{\uparrow}(s_i), & \mbox{if $(s_i,s_{i+1})$ is upward}. \\
\end{cases}
\]
Since we focus on the random binary-tree model,
we need to determine the transition probabilities where
every Dyck path is generated equiprobably.

We define a {\it monotonic path} from $(x,y)\in\domain$ as
a sequence of points on $\domain$ from $(x,y)$ to $(n-1,n-1)$
where each elementary step is either rightward or upward.
Clearly, the length of a monotonic path from $(x,y)$ is
$2(n-1)-(x+y)$, and
a monotonic path from $(0,0)$ is identical with a Dyck path.
The total number $N(x,y)$ of the monotonic paths from $(x,y)$ is written as
\begin{align}
N(x,y)&=\binom{2(n-1)-(x+y)}{n-x-1}-\binom{2(n-1)-(x+y)}{n-x-2} \nonumber \\
	&=\frac{\{2(n-1)-(x+y)\}!}{(n-1-x)!(n-y)!}(x-y+1).
\label{eq:N(x,y)}
\end{align}
For the calculation of Eq. \eqref{eq:N(x,y)},
we employed the reflection principle 
familiar in random-walk theory \cite{Karr}.

There are several remarks on $N(x,y)$:
\begin{enumerate}
\item For any $(x,y)\in\domain$, $N(x,y)$ is positive.
\item $N(n-1, y) = 1$, when $y = 0,\cdots, n-1$.
\item If $(x,y)$ is on the diagonal [i.e., $(x,y)=(k,k)$], then
	$N(k,k)=\frac{\{2(n-k-1)\}!}{(n-1-k)!(n-k)!}$,
	which is known as the $(n-k-1)$th Catalan number \cite{Conway}.
\item 
	The number of Dyck paths [which can be expressed as $N(0,0)$]
	is given by the $(n-1)$th Catalan number.
	This is well-known result, going back to Cayley \cite{Cayley}.
\item $N(x,y)=N(x+1,y)+N(x,y+1)$ for all $(x,y)\in\domain$, 
where we set $N(x,y)=0$ if $(x,y)\not\in\domain$.
\end{enumerate}

On each point $(x,y)\in\domain$,
we define transition probabilities $P_{\rightarrow}(x,y)$
and $P_{\uparrow}(x,y)$ as
\begin{subequations}
\label{eq:transition}
\begin{eqnarray}
P_{\rightarrow}(x,y) &=& \frac{N(x+1,y)}{N(x,y)}
=\frac{(n-1-x)(x-y+2)}{(1+x-y)\{2(n-1)-(x+y)\}},\\
P_{\uparrow}(x,y) &=& \frac{N(x,y+1)}{N(x,y)}
=\frac{(n-y)(x-y)}{(1+x-y)\{2(n-1)-(x+y)\}}.
\end{eqnarray}
\end{subequations}
Specifically, $P_{\rightarrow}+P_{\uparrow}\equiv 1$,
$P_{\uparrow}(k,k) = 0$ and $P_{\rightarrow}(n-1, y) = 0$.
It is also proved inductively that Eqs. \eqref{eq:transition}
realize random generation of Dyck paths.

\begin{figure}
\centering
\includegraphics[width = 0.95\textwidth]{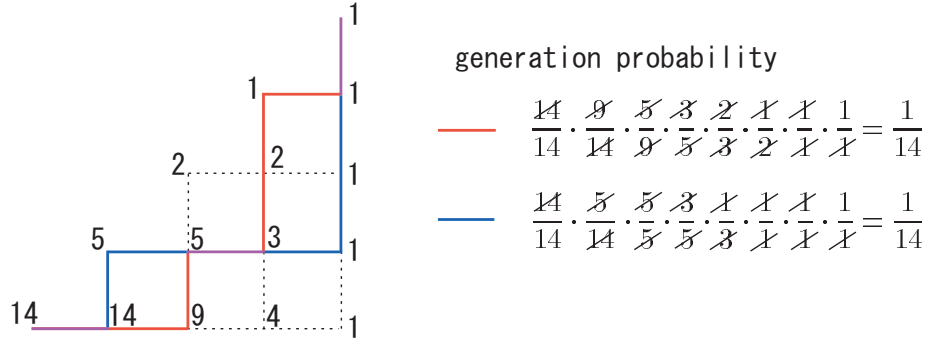}
\caption{
An example of the graphical representation of
generation probability ($n=5$).
The dashed lines indicate the grid line of $\domain$.
Each number near a lattice point indicates $N(x,y)$.
From successive canceling,
we can see that all Dyck paths are generated with the same probability.
}
\label{fig:generation}
\end{figure}

Next, we propose a graphical representation
of random Dyck paths.
The number $N(x,y)$ can be calculated graphically as follows:
\begin{enumerate}
\renewcommand{\labelenumi}{\centering(\roman{enumi})}
\item Set $N(n-1, y) = 1$ for all the rightmost points
	$(n-1, y)\; (y=0,\cdots,n-1)$ of $\domain$.
	This implies that there is only one monotonic path from $(n-1,y)$,
	which is composed only of upward steps.
\item For convenience, 
	let $N(x,y) = 0$ for all $(x,y)\not\in\domain$.
\item $N(x,y)$ is calculated from $N(x,y)=N(x+1,y)+N(x,y+1)$, that is,
	$N(x,y)$ is given by the sum of the value $N$
	on the right and upper adjacent points
	[thus, $N(x,y)$ is calculated from right to left, top to bottom].
	This implies that the monotonic paths from $(x,y)$
	consist of ones passing through $(x+1, y)$ and $(x,y+1)$.
\end{enumerate}
Note that $N(x,y)$ determined from (i)-(iii) is identical with
Eq. \eqref{eq:N(x,y)}.
The graphical representation and examples of generation probability
 is depicted in Fig. \ref{fig:generation}.
We can roughly confirm the uniformity of generated Dyck paths
through successive canceling.

\section{Numerical Procedure}
The Gaussian distribution function with mean $0$ and variance $\sigma^2$
is written as
\begin{equation}
\int_{-\infty}^x\frac{1}{\sqrt{2\pi}\sigma}
e^{-\frac{t^2}{2\sigma^2}}\,{\mathrm d}t
=\frac{1}{2}\erf\left(\frac{x}{\sqrt{2}\sigma}\right)+\frac{1}{2},
\label{eq:error}
\end{equation}
where $\erf(x)$ is the error function defined as
\[
\erf(x)\equiv\frac{2}{\sqrt{\pi}}\int_{0}^{x}e^{-t^2}\mathrm{d}t.
\]
Thus, the central limit theorems
\eqref{eq:general-1} and \eqref{eq:general-2}
are respectively rewritten as
\begin{subequations}
\label{eq:central_error}
\begin{eqnarray}
P_n\left(\sqrt{n}\left(\frac{S_{r,n}}{S_{r-1,n}}-\frac{1}{4}\right)\le x\right)
&\xrightarrow{n\to\infty}&
\frac{1}{2}\erf\left(\frac{x}{\sqrt{2}\sigma_r}\right)+\frac{1}{2}, \\
P_n\left(\sqrt{n}\left(\frac{S_{r,n}}{n}-\frac{1}{4^{r-1}}\right)\le x\right)
&\xrightarrow{n\to\infty}&
\frac{1}{2}\erf\left(\frac{x}{\sqrt{2}\tilde{\sigma}_r}\right)+\frac{1}{2}.
\end{eqnarray}
\end{subequations}

A numerical algorithm for the calculation of $\sigma_r$ and $\tilde{\sigma}_r$
is summarized as follows:
\begin{enumerate}
\renewcommand{\labelenumi}{(\roman{enumi})}
\item Generate Dyck sequences of length $2(n-1)$ randomly,
	on the basis of the method in Sec. \ref{sec:generation}.
\item Calculate Horton-Strahler indices of the Dyck sequences.
\item Compute values of both
	$\sqrt{n}\left(\frac{S_{r,n}}{S_{r-1,n}}-\frac{1}{4}\right)$ and
	$\sqrt{n}\left(\frac{S_{r,n}}{n}-\frac{1}{4^{r-1}}\right)$
	for $r=2,3,\cdots$.
\item Make distribution functions from the values, then
	determine the values of $\sigma_r$ and $\tilde{\sigma}_r$
	by fitting Eq. \eqref{eq:error} to the distribution functions.
\end{enumerate}

\section{Results of the Central Limit Theorem}
\begin{figure}[htbp]
\centering
\includegraphics[width=\textwidth]{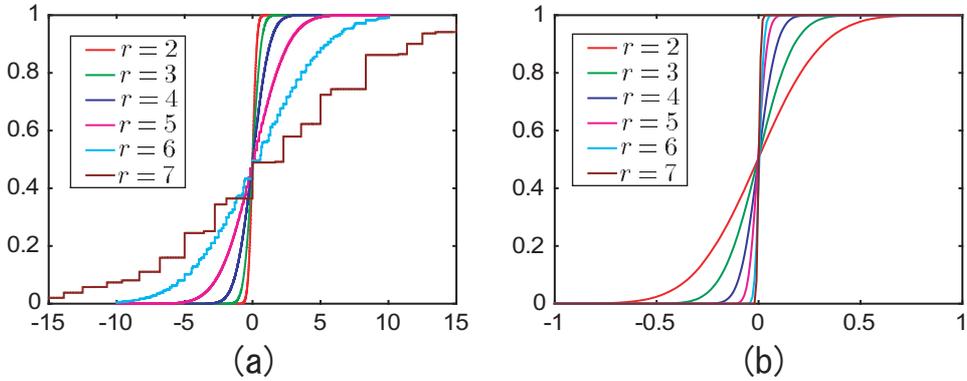}
\caption{Distribution functions of
(a) $\sqrt{n}\left(\frac{S_{r,n}}{S_{r-1,n}}-\frac{1}{4}\right)$,
and (b) $\sqrt{n}\left(\frac{S_{r,n}}{n}-\frac{1}{4^{r-1}}\right)$
with $n=10000$, $r=2-7$, generated from $10^5$ samples.}
\label{fig:distributions}
\end{figure}

Fig. \ref{fig:distributions} shows distribution functions of
$\sqrt{n}\left(\frac{S_{r,n}}{S_{r-1,n}}-\frac{1}{4}\right)$ and
$\sqrt{n}\left(\frac{S_{r,n}}{n}-\frac{1}{4^{r-1}}\right)$
generated from $10^5$ samples with $n=10000$.
The stepwise increases appear in the cases of $r=6$ and 7
in Fig. \ref{fig:distributions} (a),
because the denominator $S_{r-1,n}$ of a fraction 
$\frac{S_{r,n}}{S_{r-1,n}}$ is decreasing with respect to $r$.

\begin{figure}[htb]
\centering
\includegraphics[width=0.55\textwidth]{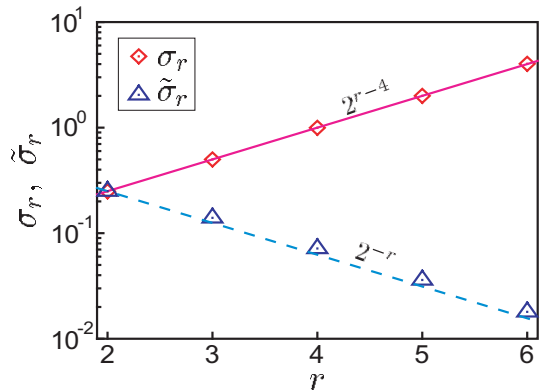}
\caption{
$r$-dependence of
$\sigma_r$ and $\tilde{\sigma}_r$.
The solid line indicates $2^{r-4}$ and
the dashed line indicates $2^{-r}$.}
\label{fig:sigmas}
\end{figure}

By fitting of the distribution function \eqref{eq:error}
to each data set in Fig. \ref{fig:distributions},
we obtain
Table \ref{tbl:sigma} and Fig. \ref{fig:sigmas}
which suggest the relations
\begin{subequations}
\label{eq:sigmas}
\begin{eqnarray}
\sigma_r &=& 2^{r-4}, \label{eq:sigma1} \\
\tilde{\sigma}_r &=& \frac{1}{2^r} \label{eq:sigma2}.
\end{eqnarray}
\end{subequations}
Eq. \eqref{eq:sigma1} is good agreement with our numerical results.
Eq. \eqref{eq:sigma2} also seems to be consistent with our results,
although there are errors of about a few percent ($\le 4\%$)
between $r$ and $-\log_2\tilde{\sigma}_r$.

\begin{table}[htb]
\centering
\caption{
Values of $\sigma_r$ and $\tilde{\sigma}_r$ obtained by fitting.
}
\begin{tabular}{c|cc|ccc}
\hline
$r$ & $\sigma_r$ & $2^{r-4}$ & $\tilde{\sigma}_r$ & $2^{-r}$
	& $-\log_2\tilde{\sigma}_r$ \\
\hline\hline
2 & 0.2492	& 0.25	& 0.2502	& 0.25 		& 1.999 \\
3 & 0.4999	& 0.5	& 0.1398	& 0.125 	& 2.839 \\
4 & 0.9968	& 1		& 0.07165	& 0.0625 	& 3.803 \\
5 & 2.0001	& 2		& 0.03605	& 0.03125 	& 4.794 \\
6 & 4.0250	& 4		& 0.01798	& 0.015625 	& 5.797 \\
\hline
\end{tabular}
\label{tbl:sigma}
\end{table}

In conclusion, the two central limit theorems are stated as
\begin{subequations}
\label{eq:result}
\begin{eqnarray}
\sqrt{n}\left(\frac{S_{r,n}}{S_{r-1,n}}-\frac{1}{4}\right)
&\Rightarrow& N\left(0, 4^{r-4}\right), \label{eq:result1}\\
\sqrt{n}\left(\frac{S_{r,n}}{n}-\frac{1}{4^{r-1}}\right)
&\Rightarrow& N\left(0,4^{-r}\right). \label{eq:result2}
\end{eqnarray}
\end{subequations}
Note that both
Eqs. \eqref{eq:result} are reduced to Eq. \eqref{eq:central} when $r=2$.

\section{Discussion}
The Horton-Strahler index is based on `merging' or `joining'
of branches in a binary tree,
and a Dyck sequence generated from the method in Sec. \ref{sec:correspond}
preserves a merging structure of the initial binary tree.
Thus, the correspondence presented in this paper
is suitable for the calculation of Horton-Strahler indices.
It is known that there are some other ways
of one-to-one correspondence between
Dyck paths and binary trees \cite{Conway, Alonso2, Viennot}.
However, Dyck paths generated from such other methods
are not directly connected to the Horton-Strahler indices.

\begin{figure}[htb]
\centering
\includegraphics[width=0.7\textwidth]{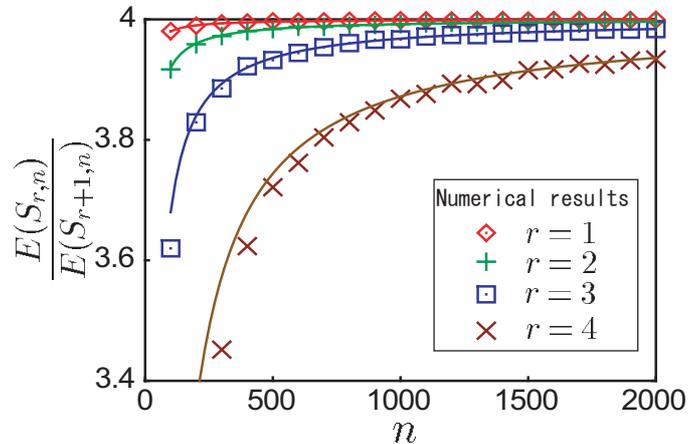}
\caption{Comparison between analytical and numerical results
of bifurcation ratios.
Points denote numerical result, and
lines denote asymptotic forms $4-\frac{4^r}{2n}$ for $r=1,2,3,4$.
Numerical data are generated from $10^5$ samples for each $n$
at intervals of 100.}
\label{fig:Moon}
\end{figure}

Our method can supply 
various numerical calculations based on the random binary-tree model,
not only the central limit theorems.
For example, see Fig. \ref{fig:Moon},
our method is able to reproduce an asymptotic expansion
of the bifurcation ratio
\begin{equation}
\frac{E(S_{r,n})}{E(S_{r+1,n})}=4-\frac{4^r}{2n}+O(n^{-2})
\quad r\ge1,
\label{eq:Moon}
\end{equation}
quite well,
which has been obtained analytically by Moon \cite{Moon}.
Moreover, for systems other than the random model,
we expect that our method is effective
with some modification of transition probabilities.

Generation of random Dyck paths can be regarded as
a Markov process on $\domain$, which is called 
the Bernoulli excursion \cite{Takacs}.
In addition, with taking a certain scaling limit,
the Bernoulli excursion converges weakly to a diffusion process
called the Brownian excursion \cite{Gikhman},
which is defined as one-dimensional Brownian motion $\{B(t):0\le t\le1\}$
such that 
$P(B(0)=0)=P(B(1)=0)=1$ and $P(B(t)\ge0)=1$ for $0\le t\le1$.
We expect that some asymptotic properties
of the random binary-tree model are derived
from the corresponding scaling limit.

Furthermore,
the number $N(x,y)$ given by Eq. \eqref{eq:N(x,y)} is 
an example of the Kostka number,
appearing in some combinatorial problems
\cite{Macdonald, Stanley}.
It is expected that
such other systems are related to a generation of random Dyck paths.

\section{Conclusion}
In the present paper,
we propose a numerical method of generating random binary trees
in the form of Dyck sequences.
We also propose a method of calculating the Horton-Strahler indices
from Dyck sequences.
From numerical results,
we confirm that the variances $\sigma_r$ and $\tilde{\sigma}_r$
are determined as Eqs. \eqref{eq:sigmas}.
Therefore,
validity of the central limit theorems
\eqref{eq:result} are suggested numerically.

\end{document}